\documentclass{ifacconf}

\usepackage{graphicx}      
\usepackage{natbib}        
\usepackage{amsmath}
\usepackage{amssymb}
\usepackage{circuitikz}
\usepackage{subcaption}
\usepackage{listings}

\begin{document}
\begin{frontmatter}

\title{{Hierarchical RL-MPC Control for Dynamic Wake Steering in Wind Farms}\thanksref{footnoteinfo}} 

\thanks[footnoteinfo]{This work has been submitted to IFAC for possible publication \\ The authors gratefully acknowledge the computational and data resources provided on the Sophia HPC Cluster at the Technical University of Denmark, DOI: 10.57940/FAFC-6M81.  \\
This study is supported by the IntelliWind project, funded through the EU MSCA Doctoral Networks programme with grant number 101168725. The support is greatly appreciated. 
}

\author[First]{Marcus Binder Nilsen} 
\author[First]{Teodor Olof Benedict Åstrand} 
\author[First]{Tuhfe Göçmen}
\author[First]{Pierre-Elouan Réthoré}

\address[First]{Technical University of Denmark, Department of Wind and Energy Systems. 
   Roskilde, Frederiksborgvej 399, Denmark (e-mail: manils@dtu.dk).}

\begin{abstract} 
Wind farm wake steering optimization is challenging due to complex flow physics and changing conditions. This paper presents a hierarchical framework that combines reinforcement learning with model predictive control, where an RL agent learns compensatory state estimates for an MPC controller, rather than directly controlling turbines. Evaluated on a three-turbine case, the approach achieves a 23\% power gain over the baseline control and surpasses the idealized MPC with perfect state knowledge. Compared to direct RL control, the hybrid architecture maintains superior safety characteristics during training while achieving comparable performance with more stable control actions.
\end{abstract}


\end{frontmatter}

\section{Introduction}

Wind energy is one of the fastest-growing renewable energy sources, driven by the global push toward carbon neutrality. Modern offshore wind farms now reach gigawatt scales, making efficient operation increasingly important for maximizing power production and reducing the levelized cost of energy \citep{wes-7-2271-2022}.

In wind farms, wakes from upstream turbines interact with downstream machines. While shared infrastructure and compact layouts offer benefits, wake interactions introduce velocity deficits and elevated turbulence that decrease downstream power and increase fatigue loads. Accurately modeling and mitigating these effects remains a central challenge in wind farm control \citep{wes-7-2271-2022}.

Yaw steering is a promising strategy to reduce wake losses by intentionally misaligning upstream turbines to deflect their wakes. This can increase total farm power despite reduced production from the yawed turbines \citep{7962923}. However, selecting optimal yaw angles is challenging because wake behavior is nonlinear, turbulent, and highly sensitive to atmospheric conditions such as wind direction, speed, and turbulence intensity.

To address these problems, two differing approaches have recently gained traction:

\begin{itemize}
    \item \textbf{Model Predictive Control (MPC):}  
    MPC leverages an internal model to optimize future control. 
    It provides interpretability and guarantees constraint compliance, but its validity depends on the accuracy of the internal model.
    
    \item \textbf{Reinforcement Learning (RL):}  
    RL enables data-driven control through direct interaction with the given environment, which could be numerical simulations or real systems.  
    It can learn policies that adapt to unmodeled effects; however, with the cost of reduced explainability and potential safety concerns.
\end{itemize},

To date, both methods have been employed separately in the literature for the task of wind farm flow control. One example of applying MPC is from the work of \citet{MPCfloridyn}, who developed the FLORIDyn model, which serves as the internal model for the MPC solver. Alternatively, for a comprehensive overview of RL-based approaches in the field, we refer to \citep{abkar2023reinforcement, Gocmen2024}.

In this work, we combine these two methods by proposing a hierarchical RL–MPC framework for wind farm control.  
The key idea is to combine a physics-based MPC controller, operating on a fast, pseudo-transient model of the wind farm, with a single RL agent.
The RL agent learns to adjust the MPC’s internal state estimates of inflow conditions, effectively compensating for the mismatch between the simplified and the true wind farm models.  
This approach exploits the strengths of both methods. Namely, the structure and explainability of MPC, as well as the adaptability and model-free learning of RL.


\section{Methodology}
\label{sec:methodology}

\subsection{Wind Farm Simulation Environment}

\begin{figure}
    \centering
    \includegraphics[width=0.95\linewidth]{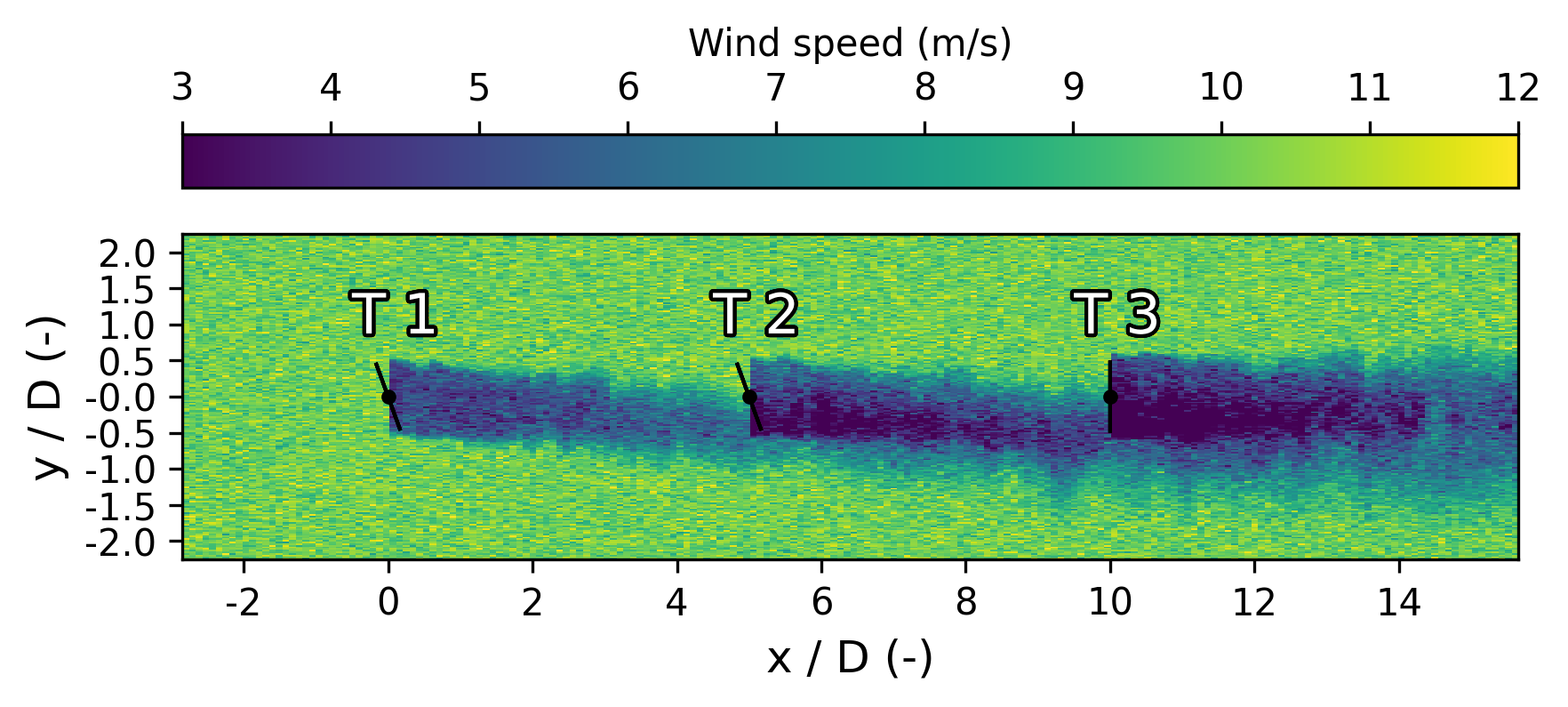}
    \caption{Visualization of the wind farm (environment) with a westerly inflow state}
    \label{fig:flow_pic}
\end{figure}

This work utilizes WindGym \citep{WindGym} as the simulation environment, serving as a surrogate for a real wind farm. WindGym provides a medium-fidelity representation of wind farm dynamics while remaining computationally feasible for training RL. It is built on DYNAMIKS \citep{Dynamiks} and uses the dynamic wake meandering model \citep{larsen2007dynamic} to simulate the transient wake dynamics. The simulated wind farm consists of $N_T = 3$ turbines arranged in a $3\times1$ row configuration with $5D$ spacing (where $D$ is the rotor diameter). An illustration of the flow field is provided in Figure \ref{fig:flow_pic}.

The environment is configured with the following parameters:
\begin{itemize}
    \item \textbf{Turbine model:} DTU10MW reference turbine \citep{DTU10MW}.
    \item \textbf{Wind conditions:} Wind speed is constant at 10 m/s, wind direction varies between $260^\circ$--$280^\circ$, and turbulence intensity is fixed at 7\%.
    \item \textbf{Temporal resolution:} The simulation time step is 5 seconds.
    \item \textbf{Yaw constraints:} Maximum yaw rate is $0.5^\circ$/s.
\end{itemize}

\subsection{Overall Framework Architecture}

\begin{figure}[htbp]
    \centering
    
    \begin{subfigure}[b]{0.8\linewidth}
        \centering
        \resizebox{\linewidth}{!}{%
        \tikzset{%
        wind turbine/.pic={
          \tikzset{path/.style={fill, draw=white, ultra thick, line join=round}}
          \path [path] 
            (-.25,0) arc (180:360:.25 and .0625) -- (.0625,3) -- (-.0625,3) -- cycle;
          \foreach \i in {90, 210, 330}{
            \path [path, shift=(90:3), rotate=\i] 
              (.5,-.1875) arc (270:90:.5 and .1875) arc (90:-90:1.5 and .1875);
          }
          \path [path] (0,3) circle [radius=.25];
        }}
        
        \begin{tikzpicture}
            \node[shape=rectangle, draw, line width=1pt, minimum width=3.59cm, minimum height=1.8cm] at (11.063, 9){} node[anchor=center, align=center, text width=3.202cm, inner sep=6pt, font=\Large] at (11.063, 9){RL};
            \path (9.5, 5.8) pic[scale=0.35] {wind turbine=1};
            \path (11.063, 5.8) pic[scale=0.35] {wind turbine=1};
            \path (12.626, 5.8) pic[scale=0.35] {wind turbine=1};
            \draw[line width=3pt, -latex] (12.875, 9) -| (16, 6.0) |- (12.875, 6.0); 
            \draw[line width=3pt, -latex] (9.25, 6) |- (5.813, 6.0) -- (5.813, 8.813) |- (9.25, 9);
            \node[shape=rectangle, minimum width=0.715cm, minimum height=0.715cm, font=\huge](N1) at (15.5, 7.5){} node[anchor=center] at (N1.text){\huge{${\gamma}$}};
            \node[shape=rectangle, minimum width=2.465cm, minimum height=1.465cm, font=\huge](N2) at (6.8, 7.5){} node[anchor=center] at (N2.text){\huge{$s_{\text{env}}$}};
        \end{tikzpicture}
        }
        \caption{RL-based control}
        \label{fig:rl_only}
    \end{subfigure}
    
    \vspace{1em}
    
    \begin{subfigure}[b]{0.8\linewidth}
        \centering
        \resizebox{\linewidth}{!}{%
        \tikzset{%
        wind turbine/.pic={
          \tikzset{path/.style={fill, draw=white, ultra thick, line join=round}}
          \path [path] 
            (-.25,0) arc (180:360:.25 and .0625) -- (.0625,3) -- (-.0625,3) -- cycle;
          \foreach \i in {90, 210, 330}{
            \path [path, shift=(90:3), rotate=\i] 
              (.5,-.1875) arc (270:90:.5 and .1875) arc (90:-90:1.5 and .1875);
          }
          \path [path] (0,3) circle [radius=.25];
        }}
        
        \begin{tikzpicture}
            \node[shape=rectangle, draw, line width=1pt, minimum width=4.6cm, minimum height=2.715cm] at (22.813, 7.25){} node[anchor=north, align=center, text width=3.202cm, inner sep=6pt, font=\Large] at (22.813, 8.625){MPC};
            \node[shape=rectangle, draw, line width=1pt, minimum width=3.0cm, minimum height=0.715cm] at (22.75, 7.3){} node[anchor=center, align=center, text width=2.6cm, inner sep=6pt, font=\large] at (22.75, 7.3){Model};
            \node[shape=rectangle, draw, line width=1pt, minimum width=3.0cm, minimum height=0.715cm] at (22.75, 6.3){} node[anchor=center, align=center, text width=2.6cm, inner sep=6pt, font=\large] at (22.75, 6.3){Optimization};
            \draw[-latex, line width=1.5pt] (24.25, 7.3) -| (24.7, 6.3) |- (24.25, 6.3);
            \draw[-latex, line width=1.5pt] (21.25, 6.3) -| (20.8, 7.3) |- (21.25, 7.3);
            \path (21.2, 3.8) pic[scale=0.35] {wind turbine=1};
            \path (22.781, 3.8) pic[scale=0.35] {wind turbine=1};
            \path (24.362, 3.8) pic[scale=0.35] {wind turbine=1};
            \draw[line width=3pt, -latex] (25.1, 7.25) -| (27.5, 4.0) |- (24.562, 4.0); 
            \draw[line width=3pt, -latex] (20.937, 4.0) |- (17.5, 4.0) -- (17.5, 7.125) |- (20.5, 7.25);
            \node[shape=rectangle, minimum width=0.715cm, minimum height=0.715cm, font=\huge](N1) at (27.0, 5.5){} node[anchor=center] at (N1.text){\huge{${\gamma}$}};
            \node[shape=rectangle, minimum width=2.465cm, minimum height=1.465cm, font=\huge](N2) at (18.5, 5.5){} node[anchor=center] at (N2.text){\huge{$s^{\text{MPC}}$}};
        \end{tikzpicture}
        }
        \caption{MPC-based control}
        \label{fig:mpc_only}
    \end{subfigure}
    
    \vspace{1em}
    
    \begin{subfigure}[b]{0.8\linewidth}
        \centering
        \resizebox{\linewidth}{!}{%
        \tikzset{%
        wind turbine/.pic={
          \tikzset{path/.style={fill, draw=white, ultra thick, line join=round}}
          \path [path] 
            (-.25,0) arc (180:360:.25 and .0625) -- (.0625,3) -- (-.0625,3) -- cycle;
          \foreach \i in {90, 210, 330}{
            \path [path, shift=(90:3), rotate=\i] 
              (.5,-.1875) arc (270:90:.5 and .1875) arc (90:-90:1.5 and .1875);
          }
          \path [path] (0,3) circle [radius=.25];
        }}
        
        \begin{tikzpicture}
            \node[shape=rectangle, draw, line width=1pt, minimum width=2.5cm, minimum height=1.8cm] at (6.35, 7.375){} node[anchor=center, align=center, text width=3.202cm, inner sep=6pt, font=\Large] at (6.35, 7.375){RL};
            \node[shape=rectangle, draw, line width=1pt, minimum width=4.6cm, minimum height=2.715cm] at (12.063, 7.375){} node[anchor=north, align=center, text width=3.202cm, inner sep=6pt, font=\Large] at (12.063, 8.75){MPC};
            \node[shape=rectangle, draw, line width=1pt, minimum width=3.0cm, minimum height=0.715cm] at (12, 7.4){} node[anchor=center, align=center, text width=2.6cm, inner sep=6pt, font=\large] at (12, 7.4){Model};
            \node[shape=rectangle, draw, line width=1pt, minimum width=3.0cm, minimum height=0.715cm] at (12, 6.5){} node[anchor=center, align=center, text width=2.6cm, inner sep=6pt, font=\large] at (12, 6.5){Optimization};
            \draw[-latex, line width=1.5pt] (13.5, 7.4) -| (13.95, 6.5) |- (13.5, 6.5);
            \draw[-latex, line width=1.5pt] (10.5, 6.5) -| (10.05, 7.4) |- (10.5, 7.4);
            \draw[line width=3pt, -latex] (7.59, 7.375) -- (9.78, 7.375);
            \path (7.95, 3.9) pic[scale=0.35] {wind turbine=1};
            \path (9.5, 3.9) pic[scale=0.35] {wind turbine=1};
            \path (11.05, 3.9) pic[scale=0.35] {wind turbine=1};
            \draw[line width=3pt, -latex] (14.35, 7.375) -| (14.75, 4.1) |- (11.25, 4.1);
            \draw[line width=3pt, -latex] (7.75, 4.1) |- (4.25, 4.1) -- (4.25, 7.25) |- (5.125, 7.375);
            \node[shape=rectangle, minimum width=0.715cm, minimum height=0.715cm, font=\huge](N1) at (14.2, 5.5){} node[anchor=center] at (N1.text){\huge{${\gamma}$}};
            \node[shape=rectangle, minimum width=2.465cm, minimum height=1.465cm, font=\huge](N2) at (8.5, 8.05){} node[anchor=center] at (N2.text){\huge{$\hat{s}^{\text{MPC}}$}};
            \node[shape=rectangle, minimum width=2.465cm, minimum height=1.465cm, font=\huge](N3) at (5.0, 5.5){} node[anchor=center] at (N3.text){\huge{$s_{\text{env}}$}};
        \end{tikzpicture}
        }
        \caption{Hierarchical RL-MPC}
        \label{fig:MPCRL}
    \end{subfigure}
    
    \caption{Comparison of wind farm control architectures: (a) Pure RL directly controlling yaw angles, (b) Pure MPC relying on accurate state estimates, (c) Hierarchical RL-MPC combining adaptive state estimation with physics-based optimization.}
    \label{fig:control_comparison}
\end{figure}
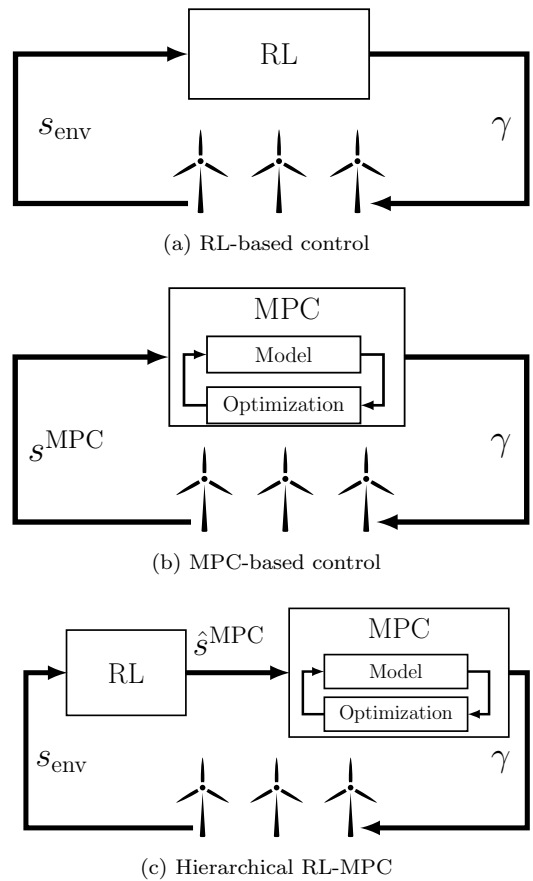

Before presenting our proposed approach, we first establish the landscape of control architectures for wind farm optimization. Figure~\ref{fig:control_comparison} illustrates three fundamental types:

\textbf{Pure RL Control (Fig.~\ref{fig:rl_only}):} The RL agent directly maps environmental observations $s_{\text{env}}$ to yaw commands $\gamma$, learning optimal control through trial and error. While this approach can adapt to complex dynamics, it requires extensive exploration and provides limited interpretability or safety guarantees.

\textbf{Pure MPC Control (Fig.~\ref{fig:mpc_only}):} The MPC controller relies on an internal model to compute optimal yaw trajectories based on estimated state $s^{\text{MPC}}$. This approach provides constraint satisfaction and interpretability, but it depends on accurate state estimation and the fidelity of the internal model.



\textbf{Hierarchical RL–MPC Framework (Fig.~\ref{fig:MPCRL}):}
The proposed control architecture is a two-level hierarchy that combines RL with MPC. At each control step \(t\), the wind farm provides local turbine measurements \(s_{\mathrm{env},t}\), which define the RL agent's observation. A Soft Actor–Critic (SAC) \citep{SACpaper} policy \(\pi_{\phi}\) then maps these observations to actions. The action of the RL agent is the estimated global state tuple, used as input for the MPC controller:
\[
a_t = \pi_{\phi}(s_{\mathrm{env},t})=s^{\mathrm{MPC}}_t = (\hat{U}_\infty,\ \hat{\mathrm{wd}},\ \hat{\mathrm{TI}})
\]
This is the inferred inflow wind speed, direction, and turbulence intensity. 

Given this estimated state, the MPC solves a constrained receding-horizon yaw optimization problem using an explicit pseudo-dynamic wake model:
\[
\boldsymbol{\gamma}^* = \arg\max_{\boldsymbol{\gamma}} 
J_{\mathrm{MPC}}\!\left(\boldsymbol{\gamma}
\mid s^{\mathrm{MPC}}_t,\, \boldsymbol{\gamma}(t)\right),
\]
Here $\boldsymbol{\gamma}$ is a vector containing the yaw angles of the turbines, subject to actuator constraints of $0.5$ deg/s. See section \ref{sec:MPC} for further implementation details regarding the MPC formulation.


This separation of roles enables the RL agent to focus solely on learning global environmental patterns from local measurements, while the MPC ensures that physically feasible and safe yaw adjustments are made. The full loop runs with a sampling time of \(60\)~s.



\subsection{Model Predictive Control Formulation}
\label{sec:MPC}

The MPC component of this work builds upon the approach of \citet{MPCfloridyn}. We implement their back-to-front optimization strategy, making key modifications for improved computational efficiency and online learning compatibility for our scenario.

\subsubsection{Pseudo-Dynamic Wake Model in PyWake}

Instead of using FLORIDyn, our MPC controller relies on a custom pseudo-dynamic wake model built on PyWake \citep{pywake}, that explicitly accounts for wake advection delays. 

\textbf{Delay modeling:} The advection delay between upstream turbine $j$ and downstream turbine $i$ is calculated based on the projected distance:
\begin{equation}
\Delta_{j,i} = \frac{(x_i - x_j)\cos(\mathrm{wd}) + (y_i - y_j)\sin(\mathrm{wd})}{U_{\infty}}
\end{equation}
where $(x_i,y_i)$ are turbine positions, $\mathrm{wd}$ is the wind direction, and $U_{\infty}$ is the free-stream wind speed. The delay matrix $\mathbf{\Delta}\in\mathbb{R}^{N_T\times N_T}$ collects all pairwise delays; entries where $j$ is not upstream of $i$ are zero.

\textbf{Dynamic power calculation:} When computing the power of turbine $i$ at prediction step $k$, we use the yaw commands from upstream turbines at the specific past times when those wakes were generated. The effective yaw of upstream turbine $j$ as seen by turbine $i$ at $t_k$ is:
\[
\gamma^{\mathrm{eff}}_{j\to i}[k] = \gamma_j[k - d_{j,i}],
\]
where $d_{j,i}$ corresponds to the discrete time steps derived from $\Delta_{j,i}$.
PyWake is then called using the vector of effective yaw angles:
\[
P_i[k] = \text{PyWake}\!\left( \boldsymbol{\gamma}^{\mathrm{eff}}_i[k], U_\infty,\text{wd}, \text{TI} \right).
\]

We incorporate the yaw loss penalty $\omega(\gamma)$ based on \citet{MPCfloridyn}, such that the penalized power is $\bar{P_i} = P_i \cdot \omega(\gamma)$, with:
\begin{equation}
    \omega(\gamma)=\sigma(\gamma - \gamma_{\max}) \cdot \sigma(\gamma_{\min} - \gamma ) 
\end{equation}
where $\sigma(x)=0.5 \tanh(50x)+0.5$. This function creates smooth transitions at $\gamma_{\max} = 33^\circ$, effectively discouraging large yaw angles before the yaw limits are reached.

\subsubsection{Cost Function Design}
\label{subsubsec:cost}

Following \citet{MPCfloridyn}, we employ their standard energy maximization formulation without time-shifting and their use of basis functions for the yaw trajectories. This means that the yaw trajectory for turbine $i$ can be expressed as:
\begin{equation}
    \gamma_i(t)=\gamma_i(0)+\psi(o_{i,1},o_{i,2},t) 
\end{equation}
where $\mathbf{O}_i=(o_{i,1},o_{i,2})\in [0,1]$ are the optimization parameters, and the basis function is:

\begin{multline}
    \psi(o_{1},o_{2},t) = 2(o_1-0.5) \cdot \text{sat}_{[0,1]} \\ 
    \left( \dfrac{t/t_{ah}-t_{s,n}(o_2)}{2|o_1-0.5|} \right) \cdot r_{\gamma,max} \cdot t_{ah}
\end{multline}

with $t_{s,n}$ being the starting time function:
\begin{equation}
    t_{s,n}(o_2)=o_2 (1-2 | o_1-0.5|)
\end{equation}
And the situation function:
\begin{equation}
    \text{sat}_{[a,b]}(x) = \begin{cases}
        a \text{ if } x < a \\
        x \text{ if } a \leq x \leq b \\
        b \text{ if } x > b 
    \end{cases}
\end{equation}

Here $t_{ah}$ is set to 100 seconds, and the maximum yaw rate $r_{\gamma, \textbf{max}}$ is 0.5 deg/s. The basis function makes it possible to describe each yaw trajectory as a function of only 2 parameters. This turns the cost function into: 

\begin{equation}
\begin{aligned}
\max_{\mathbf{O}} \quad & J_{\text{MPC}}(\mathbf{O}) = \Delta t \sum_{i=1}^{N_T} \sum_{k=1}^{\tau_{\text{ph}}} P_i(\mathbf{O}, k) \\
\text{subject to} \quad & \boldsymbol{\gamma}[k+1] = \boldsymbol{\gamma}[k] + \boldsymbol{\psi}(\mathbf{O}, k) \\
& |\dot{\gamma}_i[k]| \leq r_{\gamma,\max} \quad \forall i, k
\end{aligned}
\end{equation}
where $\mathbf{O} \in \mathbb{R}^{2N_T}$ are the optimization parameters and $\tau_{\text{ph}}$ is the prediction horizon, set to 500 seconds in this work.

\subsubsection{Sequential Optimization with DIRECT}
\label{sec:opt}

We also adopt the back-to-front sequential optimization from \citet{MPCfloridyn}, where turbines are sorted by their position along the wind direction and optimized from downstream to upstream. 
The key modification is our use of the DIRECT optimizer implemented in SciPy \cite{2020SciPy-NMeth} instead of particle swarm optimization. 

This optimizer was chosen based on a small internal benchmark, which weighed the power gain against the computational time. The direct optimizer was found to be the fastest of the tested optimizers, and for this simple case, it consistently produced good solutions. 


\subsection{Integration with RL framework}

One main difference from the work of \citet{MPCfloridyn} is that we replace their ensemble Kalman filter directly with RL. 
This eliminates the need for complex state estimation while enabling the system to learn compensatory adjustments. For instance, if the wake model consistently underestimates wake losses, the RL agent can learn to provide adjusted wind speed estimates that result in better control performance.

\subsection{Reinforcement Learning Formulation}

The RL agent estimates global wind conditions from local turbine measurements to inform the MPC controller. We use the SAC algorithm \cite{SACpaper} implemented from \cite{huang2022cleanrl}. The Markov Decision Process (MDP) is defined as: 

\textbf{State space:} $\mathcal{S} \subset \mathbb{R}^{3N_T}$ contains normalized observations for all $N_T$ turbines:
\begin{equation}
s_t = s_{\text{env}} = [\underbrace{u_1^{\text{norm}}, \phi_1^{\text{norm}}, \gamma_1^{\text{norm}}, }_{\text{turbine 1}}, \ldots, \underbrace{u_{N_T}^{\text{norm}}, \phi_{N_T}^{\text{norm}}, \gamma_{N_T}^{\text{norm}} }_{\text{turbine } N_T}]^\top
\end{equation}
where $u$ denotes the measured wind speed, $\phi$ the measured wind direction, and $\gamma$ the current yaw offset, all normalized to $[-1, 1]$.

\textbf{Action space:} $\mathcal{A} \subset \mathbb{R}^3$ represents the estimated global wind conditions provided to the MPC:
\begin{equation}
a_t = [\hat{\mathrm{wd}}^{\text{norm}}, \hat{U}_\infty^{\text{norm}}, \hat{\mathrm{TI}}^{\text{norm}}]^\top = s_{\text{MPC}} \in [-1, 1]^3
\end{equation}
These are denormalized to physical units using linear scaling: $\hat{x} = \frac{(a_x + 1)}{2}(x_{\max} - x_{\min}) + x_{\min}$.

\textbf{Transition dynamics:} $\mathcal{P}(s_{t+1} | s_t, a_t)$ are implicitly defined by the WindGym environment and MPC controller interaction.

\textbf{Reward function:} The reward is the normalized farm power production:
\begin{equation}
r_t = \frac{ \sum_{i=1}^{N_{T}} P_{i}(t)}{N_{T} P_{\text{rated}}(\bar{u})} 
\end{equation}
where $P_{\text{rated}}(\bar{u})$ is the rated power production at the mean ambient wind speed $\bar{u}$. The discount factor is $\gamma_{\text{RL}} = 0.99$.


\subsubsection{Soft Actor-Critic Algorithm}

This work uses the SAC algorithm \cite{SACpaper}, which augments the standard RL objective with an entropy regularization term to encourage exploration:
\begin{equation}
J_{\text{SAC}}(\pi) = \mathbb{E}_{\tau \sim \pi}\left[\sum_{t=0}^{\infty} \gamma_{\text{RL}}^t (r_t + \alpha \mathcal{H}(\pi(\cdot|s_t)))\right]
\end{equation}
where $\mathcal{H}(\pi(\cdot|s_t)) = -\mathbb{E}_{a \sim \pi}[\log \pi(a|s_t)]$ is the policy entropy and $\alpha$ is the temperature parameter controlling the exploration-exploitation tradeoff. The algorithm is implemented from CleanRL \citep{huang2022cleanrl}. 
Both the actor and critic use fully connected networks (two hidden layers of 256 units with ReLU activation). 
Training proceeds with a replay buffer of size $10^6$, batch size 256, and learning rates $\eta = 3 \times 10^{-4}$. The agent is trained for $500,000$ total time steps across 30 parallel environments.

\subsection{Benchmarks}

To evaluate the proposed hierarchical framework, we compare it against three baselines:

\textbf{1. Greedy controller:} A realistic baseline, where all turbines in the farm are only optimizing their own power production, resulting in zero yaw offset.

\textbf{2. Idealized MPC:} A pure MPC controller that receives the true global inflow conditions $(\mathrm{wd}, U_\infty, \mathrm{TI})$ directly from WindGym. This shows the MPC performance assuming perfect state estimation.

\textbf{3. Direct RL (SAC-RL):} A baseline where the SAC agent directly controls turbine yaw changes without the intermediate MPC layer. This uses the same formulation as for the hierarchical approach, with the only changes being: 
\begin{itemize}
    \item \textbf{Action space:} $a_t \in [-1, 1]^{N_T}$, representing normalized yaw rate commands.
    \item \textbf{Execution:} $\Delta\gamma_i = a_i^{\text{norm}} \cdot r_{\gamma,\max} \cdot \Delta t_{\text{env}}$.
    \item \textbf{Frequency:} The Direct RL agent operates at $\Delta t_{\text{env}} = 10 \ s$, compared to the 60-second interval of the Hierarchical MPC.
\end{itemize}

\subsection{Safety Monitoring Metrics}

To assess the safety implications of different control approaches, we monitor the frequency of yaw angles exceeding $30^\circ$ over a rolling window of 1000 environment steps. This metric $V_{30}(t)$ represents the percentage of turbine-steps with what is deemed as unnecessarily large yaw offsets:

\begin{equation}
V_{30}(t) = \frac{100}{1000 \cdot N_T} \sum_{k=t-999}^{t} \sum_{j=1}^{N_T} \mathbf{1}[|\gamma_j(k)| > 30^\circ]
\end{equation}

where $\mathbf{1}[\cdot]$ is the indicator function. This metric provides insight into how aggressively each approach explores the action space, serving as a proxy for potential mechanical stress caused by the large yaw offsets.

\section{Results}
\label{sec:results}

Both the SAC-MPC and SAC-RL controllers were trained using three different random seeds to ensure statistical robustness. During training, they were evaluated every 25,000 training steps within WindGym across three inflow conditions ($265^\circ$, $270^\circ$, $ 275^\circ$) with a constant wind speed of 10 m/s and a turbulence intensity of 7\%. Evaluations were performed over 1000 simulated seconds. These methods are compared against the Greedy baseline and the Idealized MPC.

\begin{figure}
    \centering
    \includegraphics[width=1\linewidth]{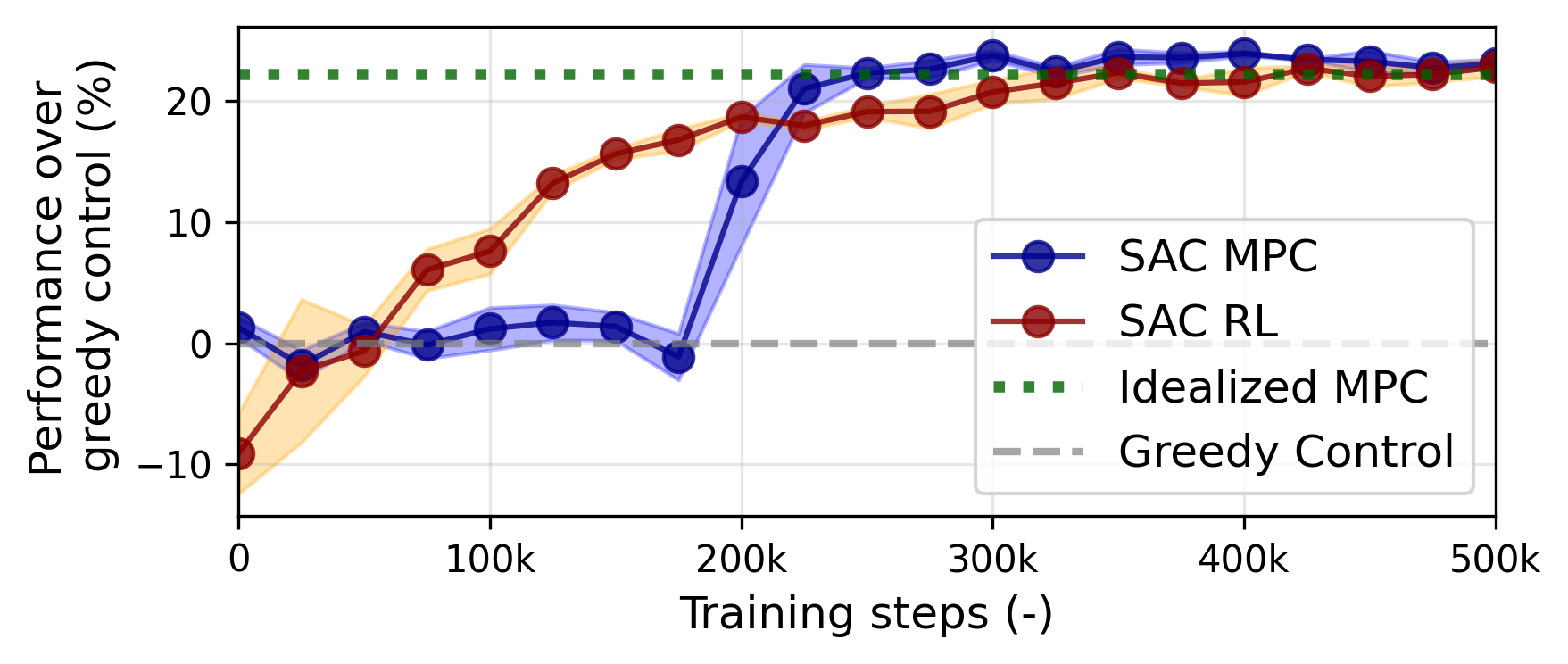}
    \caption{Plot depicting the mean farm power produced by the SAC-MPC approach throughout the model training steps, compared to the Idealized MPC and greedy baseline controller.}
    \label{fig:power_comp}
\end{figure}

Figure \ref{fig:power_comp} presents the mean power production across the three inflow conditions throughout training. The two learning-based approaches exhibit noticeably different learning dynamics. The SAC-MPC initially performs at the greedy baseline level before rapidly discovering effective control strategies around 200,000 training steps, resulting in a rapid increase from approximately 0\% to +23\% power gain. In contrast, SAC-RL begins with a 10\% power deficit before steadily improving, reaching greedy baseline performance after approximately 25,000 steps and continuing to increase thereafter to a level of 22\% power gain.

Both methods ultimately converge to similar performance levels, with SAC-MPC maintaining a slight advantage. Notably, by approximately 300,000 training steps, SAC-MPC surpasses the Idealized MPC baseline despite the latter having access to perfect state information (22.2\%). This result demonstrates that the RL agent learns to provide the MPC with compensatory estimates that account for model-environment mismatch, enabling more effective exploitation of wake interactions than would be possible with true but static inflow conditions alone.

\begin{figure}
    \centering
    \includegraphics[width=1\linewidth]{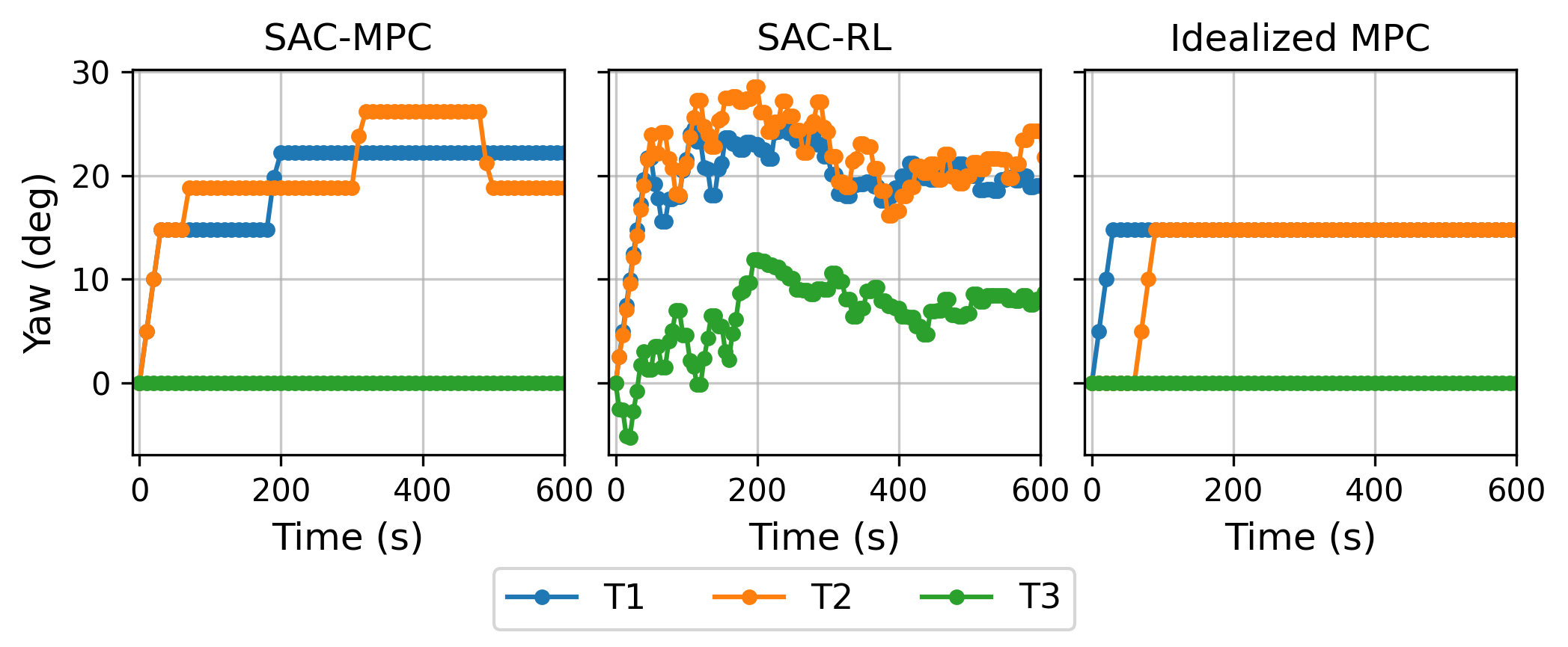}
    \caption{Yaw offset control actions, $\gamma$, for the RL and Idealized MPC controllers for a fixed wind direction of $270^\circ$.}

    \label{fig:yaw_angles}
\end{figure}

Figure \ref{fig:yaw_angles} shows the yaw angle trajectories for a single seed of the farm, being controlled by the SAC-MPC, SAC-RL, and the Idealized MPC controller for the $270^\circ$ wind direction. Note that we only show the first 500 seconds. The Idealized MPC employs a relatively conservative wake steering strategy with modest yaw offsets. In contrast, both SAC-MPC and SAC-RL implement more aggressive strategies, actively yawing the two upstream turbines to achieve stronger wake deflection.

However, important behavioral differences emerge between the two learning-based approaches. SAC-RL exhibits frequent oscillations in yaw set points. Notably, the most downstream turbine also exhibits dynamic misalignment, possibly to align better with the local inflow with meandering wake and to steer away from the partial wake effects.

\begin{figure}
    \centering
    \includegraphics[width=0.95\linewidth]{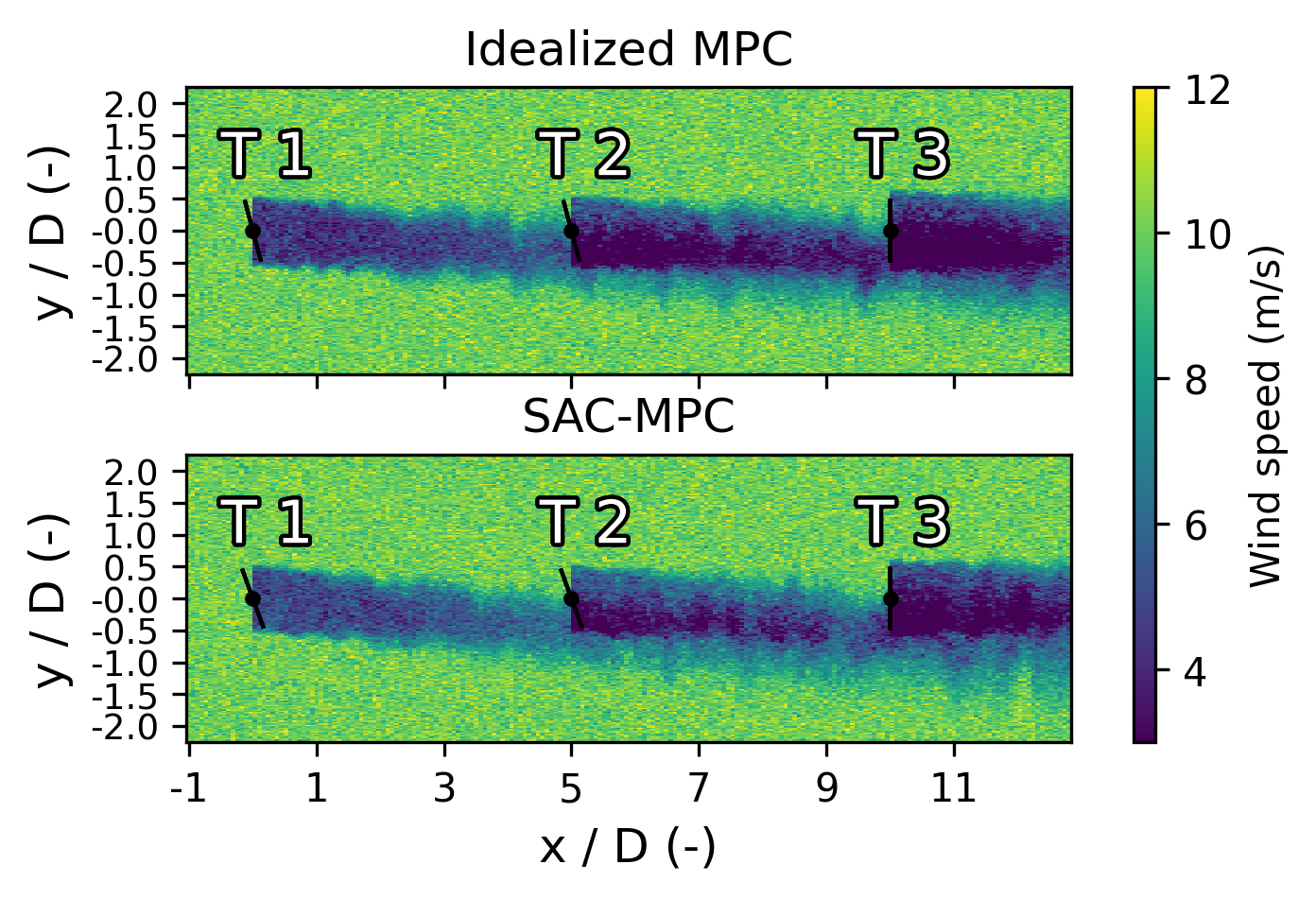}
    \caption{Flow field visualization of wake propagation under the final yaw configurations from Figure~\ref{fig:yaw_angles}. Compared to the Idealized MPC (top plot), the SAC-MPC (bottom plot) achieves stronger wake deflection. }
    \label{fig:wake_render}
\end{figure}

The resulting wake deflection is visualized in Figure \ref{fig:wake_render}, which compares flow fields under final yaw configurations for the SAC-MPC and the Idealized MPC approach. The learning-based approach achieves stronger wake deflection than the Idealized MPC, creating more favorable conditions for the downstream turbine(s) in terms of reduced wake effects and resulting in increased power gain.

\begin{figure}
    \centering
    \includegraphics[width=1\linewidth]{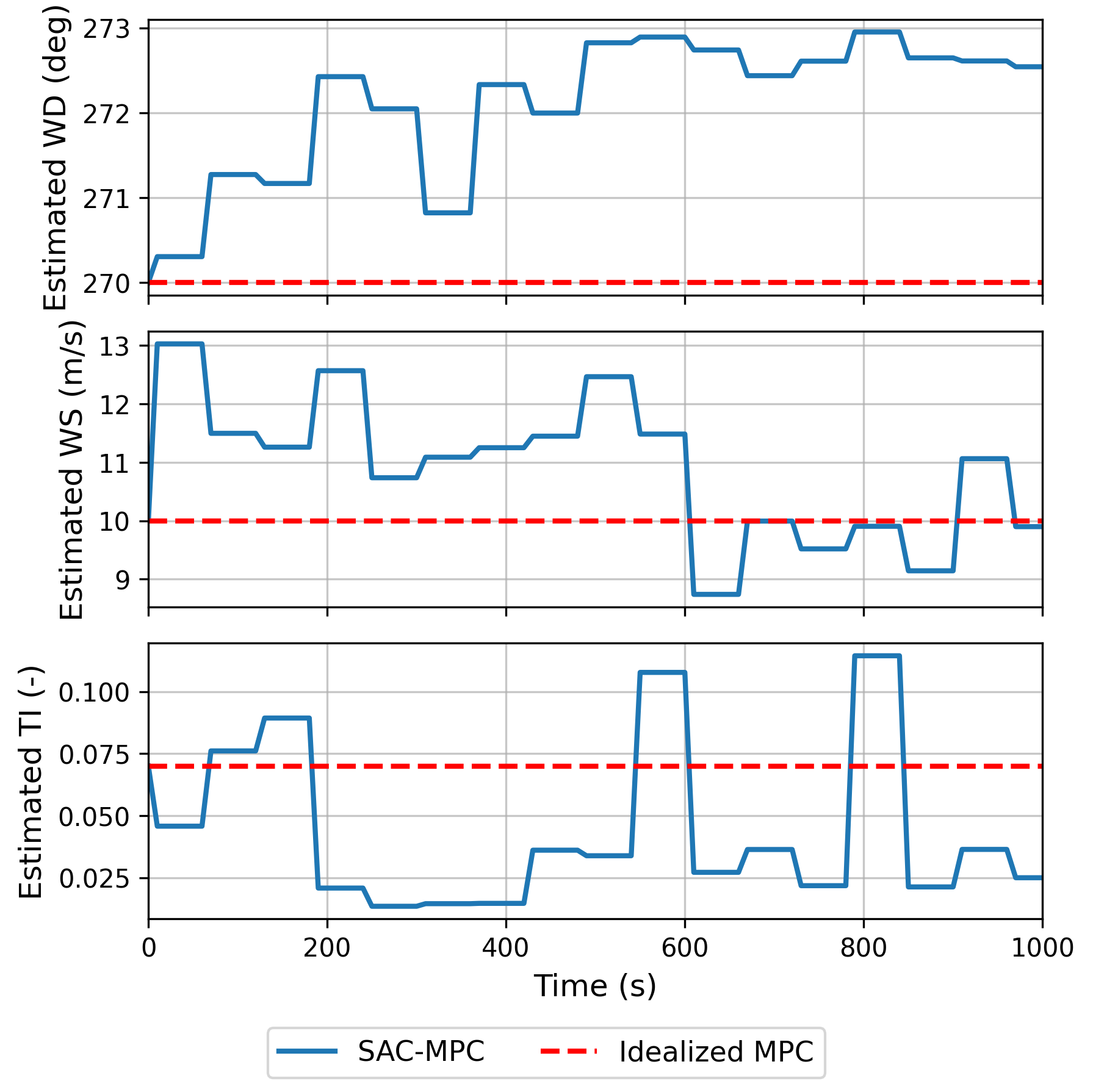}
    \caption{Plots depicting the estimated wind direction (WD), wind speed (WS), and turbulence intensity (TI) from the RL agent and the corresponding true value from WindGym fed to the MPC during the control actions in Figure \ref{fig:yaw_angles}.}
    \label{fig:estimated_inputs}
\end{figure}

Comparing the estimated inputs from the RL agent with the values used in Figure \ref{fig:estimated_inputs}, it is evident that the SAC agent provides dynamically varying estimates of the inflow, which differ from the global states (mean values over the evaluation period) employed by the Idealized MPC. This discrepancy indicates that the RL agent is not solely estimating physical states but rather learning optimal inputs that yield higher energy capture. Furthermore, the consistent offset for the wind direction and the turbulence intensity suggests that the agent compensates for a model-environment mismatch in a manner that improves the state estimations, hence the total power output.



\begin{figure}
    \centering
    \includegraphics[width=0.95\linewidth]{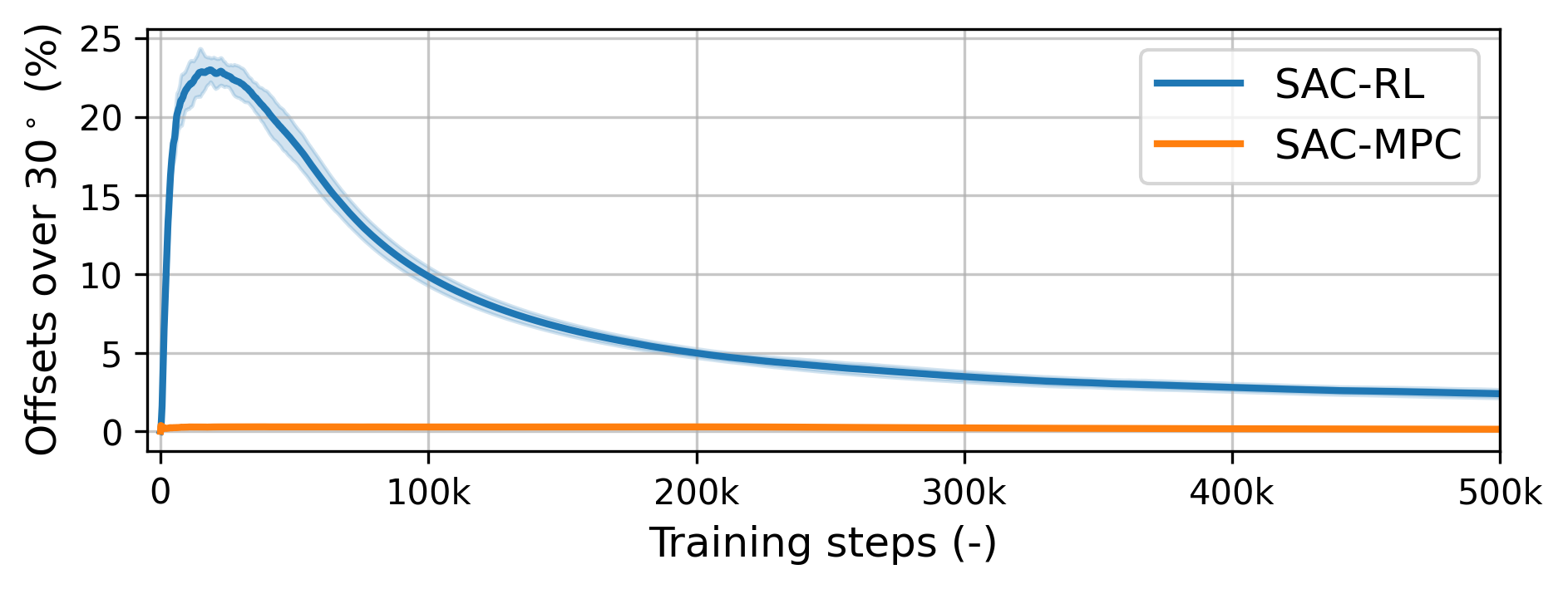}
    \caption{Percentage of time with turbines having a yaw offset of above 30 degrees}
    \label{fig:violations}
\end{figure}

Figure \ref{fig:violations} tracks the frequency of yaw angles exceeding $30^\circ$ throughout training, a metric serving as a proxy for potential mechanical stress. The SAC-RL approach exhibits a substantial amount of time exceeding 20\% during early training as it explores the action space, before gradually reducing to lower levels as the policy becomes more confident in the best actions. In contrast, the hierarchical SAC-MPC maintains consistently low levels throughout training, rarely approaching these extreme yaw angles. This difference highlights a key practical advantage of the hybrid architecture: the MPC's model-based optimization naturally bounds control actions within safe operating regions, providing implicit safety guarantees even during the exploration phase when the RL agent has not yet learned effective policies. This property could prove valuable for safe deployment in real wind farms where excessive yaw activity can reduce turbine lifespan.

In summary, the developed hierarchical RL-MPC (implemented via the SAC algorithm) is shown to improve the performance of the power optimization compared to all the benchmarks, namely Greedy controller, idealized MPC, and direct RL. The overall power gain achieved is comparable among Idealized MPC, Direct RL, and hierarchical RL-MPC, with the latter showing a slightly higher increase, while ensuring the safety compliance of the control actions even during the exploration phase.

\section{Conclusion and Future Work}
\label{sec:conclusion}

This work presents a hierarchical framework that combines reinforcement learning with model predictive control for optimizing the power output of a wind farm. The key innovation is the use of an RL agent to learn optimal state estimates for an MPC controller, enabling compensation for model-environment mismatch while maintaining the structure, interpretability, and safety properties of physics-based control.
Experimental results on a three-turbine test case demonstrate three important findings. First, the SAC-MPC framework achieves approximately 23\% power gain over greedy baseline control and surpasses even an Idealized MPC with perfect state knowledge (reaching 22.2\% increase), showing that learned compensatory estimation can outperform accurate physical measurements when processed through imperfect models. Second, while both SAC-MPC and SAC-RL achieve similar ultimate performance, the hierarchical approach maintains substantially lower levels of extreme yaw events throughout the training process, indicating lower levels of risk for field implementation. Third, SAC-MPC produces smoother control actions focused on upstream turbines, whereas SAC-RL exhibits oscillatory behavior and ineffective yawing of downstream turbines, which could increase mechanical wear. These findings demonstrate the potential of hybrid architectures such as RL-MPC  to successfully leverage complementary strengths: MPC provides structure, interpretability, and compliance to safety constraints in terms of yaw setpoints, while RL enables adaptive compensation for model inaccuracies.

There are several interesting areas for future work. First, practical wind farm control requires balancing multiple objectives beyond power maximization. Load-aware control strategies could be developed by incorporating turbine fatigue models into the MPC optimization or RL reward function, enabling multi-objective formulations that trade off power production against structural loading. Second, the framework should be validated on larger wind farm configurations with realistic layouts, extended wind conditions, and high-fidelity aeroelastic simulations to assess scalability and robustness. Finally, the learning approach could be extended beyond state estimation to adapt other model components, such as wake model hyperparameters, further improving the match between controller predictions and true system behavior.
This hierarchical RL-MPC approach opens a pathway toward wind farm controllers that are simultaneously data-driven, physics-informed, and operationally safe - addressing a critical challenge in the deployment of advanced optimization methods for real-world energy systems.

\section*{DECLARATION OF GENERATIVE AI AND AI-ASSISTED TECHNOLOGIES IN THE WRITING PROCESS}
During the preparation of this work, the authors used [ChatGPT and Claude] for coding help/debugging purposes. Furthermore, [ChatGPT, Claude, and Grammarly] have been used for grammar purposes. After using these tools, the authors reviewed and edited the content as necessary, taking full responsibility for the accuracy and content.

\vspace{-0.1cm}
\bibliography{ifacconf}             
\end{document}